\documentclass{webofc}
\usepackage[varg]{txfonts}   % Web of Conferences font
\begin{document}
\title{The HSF Conditions Database Reference Implementation}

\author{\firstname{Ruslan} \lastname{Mashinistov }\inst{1}\fnsep\thanks{\email{ruslan.mashinistov@cern.ch}}
\and
\firstname{Lino} \lastname{Gerlach}\inst{1}\fnsep\thanks{\email{lino.oscar.gerlach@cern.ch}}
\and
\firstname{Paul} \lastname{Laycock}\inst{1}\fnsep\thanks{\email{paul.james.laycock@cern.ch}}
\and
\firstname{Andrea}
\lastname{Formica}
\and
\firstname{Giacomo}
\lastname{Govi}
\and
\firstname{Chris}
\lastname{Pinkenburg}\inst{1}
}

\institute{Brookhaven National Lab (US)
}

\abstract{%
Conditions data is the subset of non-event data that is necessary to process event data. It poses a unique set of challenges, namely a heterogeneous structure and high access rates by distributed computing. The HSF Conditions Databases activity is a forum for cross-experiment discussions inviting as broad a participation as possible. It grew out of the HSF Community White Paper work to study conditions data access, where experts from ATLAS, Belle II, and CMS converged on a common language and proposed a schema that represents best practice. Following discussions with a broader community, including NP as well as HEP experiments, a core set of use cases, functionality and behaviour was defined with the aim to describe a core conditions database API. This paper will describe the reference implementation of both the conditions database service and the client which together encapsulate HSF best practice conditions data handling.

Django was chosen for the service implementation, which uses an ORM instead of the direct use of SQL for all but one method. The simple relational database schema to organise conditions data is implemented in PostgreSQL. The task of storing conditions data payloads themselves is outsourced to any POSIX-compliant filesystem, allowing for transparent relocation and redundancy. Crucially this design provides a clear separation between retrieving the metadata describing which conditions data are needed for a data processing job, and retrieving the actual payloads from storage. The service deployment using Helm on OKD will be described together with scaling tests and operations experience from the sPHENIX experiment running more than 25k cores at BNL.
}
\maketitle
\section{Introduction}
\label{sec:intro}

Conditions data refers to non-event experimental data, which characterizes the detector's status during data acquisition. This encompasses details about the detector's geometry, signal readout maps, recordings of detector control system parameters like temperature and voltage, alignment calibrations, beam attributes, and more. Conditions data is subject to ongoing changes, including actions such as recalibrating the detector using extensive cosmic datasets and refining calibration algorithms. Consequently, distinct versions of conditions data exist for different modifications made to the detector or system configurations. Conditions data are used by event processing applications that operate on a vast distributed computing infrastructure. With tens of thousands of jobs concurrently accessing the conditions data, the reading rate can reach O(10) kHz. To address this demand, the conditions database service must employ fast database queries and efficient caching mechanisms.\\
Despite the crucial role that conditions data plays in the Computing Model of any High-Energy Physics (HEP) experiment, there are no shared standards across experiments. Defining the best practices for this area remains an ongoing discussion. The HSF Conditions Database activity \cite{CondDB} serves as a central forum for these discussions, bringing together a wide range of participants from various experiments. This collaborative forum aims to establish a common vocabulary, data model, and API specification.\\
Implementing these recommendations, a reference system was developed comprising an easy-to-deploy, scalable server-side application built on PostgreSQL and a Django-REST API. On the client side, a configurable C++ library facilitates seamless interaction with the server and the handling of the payloads which reside on a distributed file system.

\section{Implementation}
\begin{figure}[h]
\centering
\includegraphics[width=\textwidth]{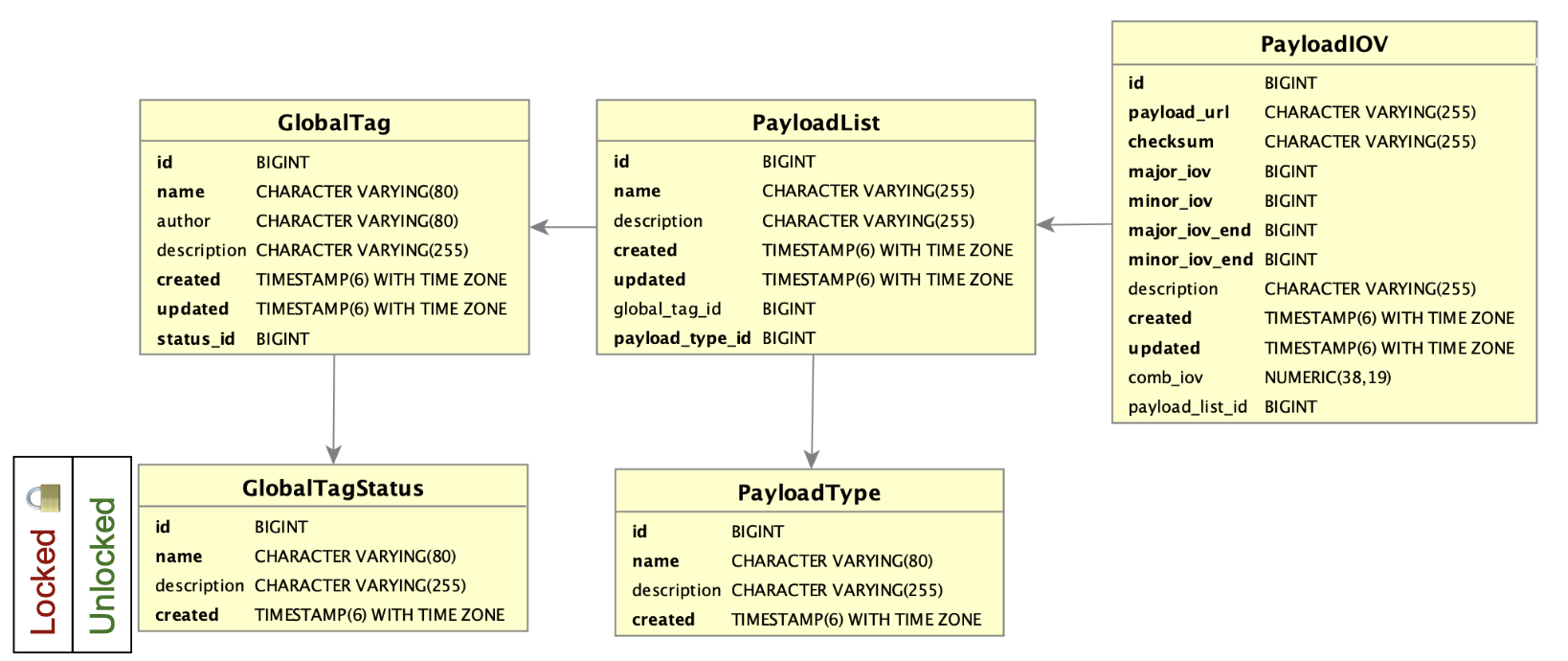}
\caption{The relational database schema of NoPayloadDB. Every PayloadIOV references an external payload and stores its metadata including the IoV. A PayloadList contains a sequence of PayloadIOVs and has one associated PayloadType. Each GlobalTag has at most one PayloadList per PayloadType and a GlobalTagStatus.}
\label{fig:schema}
\end{figure}
\subsection{Database Schema}
The server-side application is known as NoPayloadDB. It deploys a relational database schema which is based on the design proposed in the HSF White Paper \cite{laycock2019hep} and depicted in Fig. \ref{fig:schema}. The different objects in the schema will be briefly explained in the following. A \textbf{PayloadIOV} references exactly one payload on external storage via a unique identifier (\textit{payload\_url}) and contains its metadata like the file's size and checksum. Additionally, it defines the interval of validity (IoV) with a major- and minor- parameter, allowing for flexibility with respect to the granularity. For example, a collider experiment may choose run-number and luminosity block as major- and minor IoV, respectively. A \textbf{PayloadList} contains a sequence of not-overlapping PayloadIOVs and has an associated \textbf{PayloadType}, which can, for example, represent a certain subsystem of the detector. A \textbf{GlobalTag} serves as the single top-level configuration parameter. It contains at most one PayloadList for each PayloadType so that for any combination of major- and minor-IoV, a GlobalTag will resolve to at most one PayloadIOV for each PayloadType. Finally, each GlobalTag has a \textbf{GlobalTagStatus} (locked or unlocked) to protect it from unwanted write operations.\\
Django's native Object Relational Model (ORM) is used for most queries. However, the endpoint with the highest expected access rate was implemented with an optimized raw SQL query as this was found to provide a significantly better performance (see Sec. \ref{sec:testing} and Fig. \ref{fig:endpoints}). Furthermore, major- and minor IoV are combined to form a single decimal column and a covering index is defined so that the query can run as an index condition without scanning the database tables.

\subsection{Server deployment at OKD}
NoPayloadDB is a Django-based \cite{django} RESTful web service containerized using Docker \cite{docker}. It's deployed on the OKD cluster \cite{okd} of the Scientific Data and Computer Center (SDCC) \cite{sdcc} at the Brookhaven National Laboratory. OKD, an open-source container application platform, is built around Kubernetes. The deployment procedure on the OKD infrastructure is well defined and automated with a Helm chart \cite{helm} as it is shown in Fig.2. Helm serves as a Kubernetes \cite{kube} applications manager that enables the definition, installation, and upgrading of applications. Helm uses a packaging format called charts. A Helm chart is a collection of files that describe a related set of Kubernetes resources.\\
\begin{figure}[h!]
\centering
\includegraphics[width=10cm,clip]{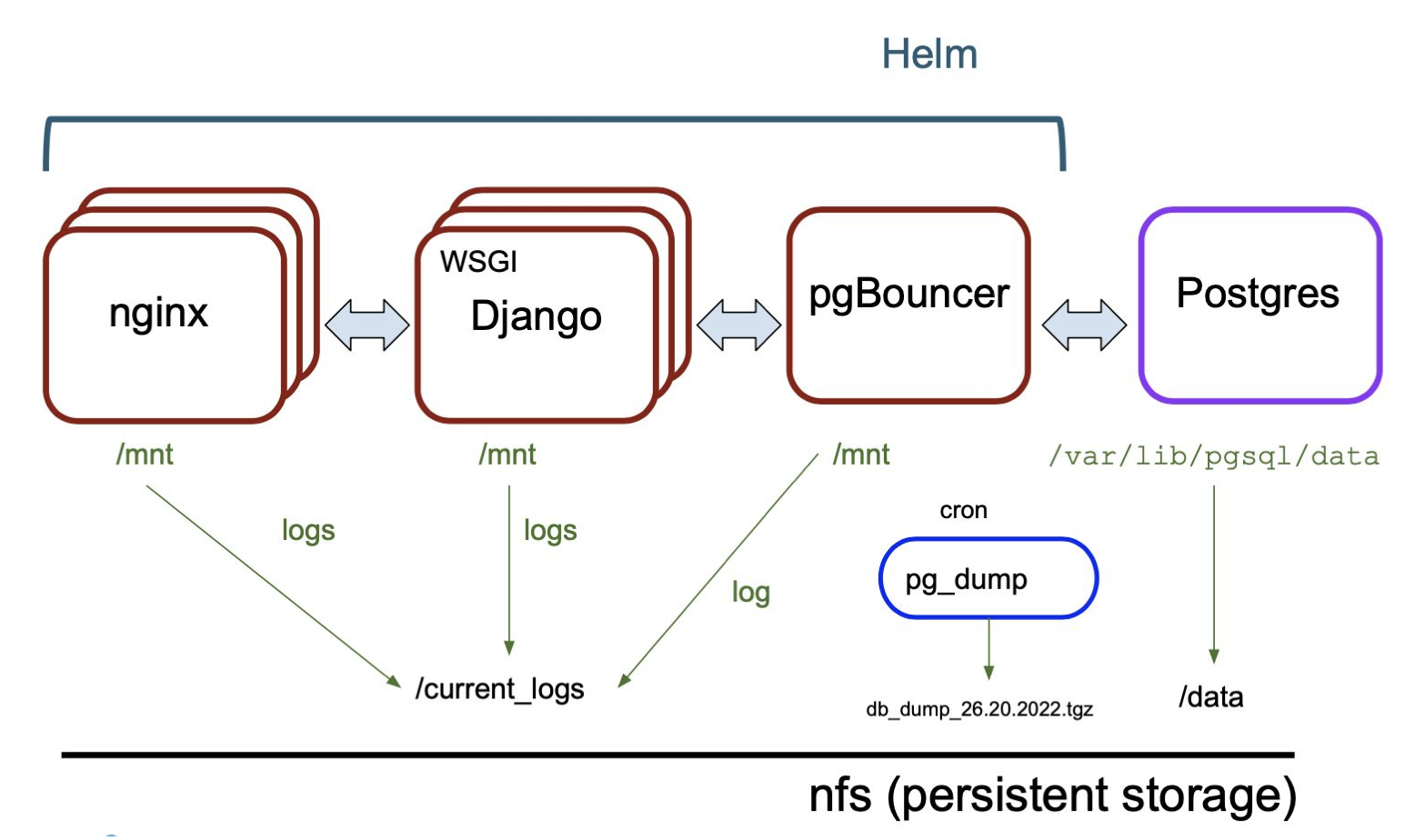}
\caption{The NoPayloadDB deployment on OKD. The Django application together with the web server, the database pooler and a Kubernetes cron job for the period database backups are included in the Helm chart.}
\end{figure}

The deployment includes the following components:
\begin{itemize}
    \item The NoPayloadDB Django application running under a gunicorn \cite{gunicorn} (WSGI) server.
    \item A Postgres \cite{postgres} database with persistent storage on NFS \cite{nfs}.
    \item An instance of pgBouncer \cite{pgbouncer} serving as a database pooler, optimizing the efficiency of database connections by keeping them open for reuse by multiple requests from Django.
    \item An Nginx \cite{nginx} web server.
    \item A Kubernetes Cron Job responsible for executing a database dump. This cron job initiates the creation of a Pod containing a PostgreSQL container, which then executes the "pg\_dump" command to generate a dump file on NFS.
\end{itemize}
The Django and nginx services can be scaled horizontally by executing them on multiple Kubernetes \textit{pods} at the same time. The default value for both is 5. Every component is configured to store its logs on the mounted NFS volume.

\subsection{Client-Side Software}
\begin{figure}[h]
\centering
\includegraphics[width=8cm,clip]{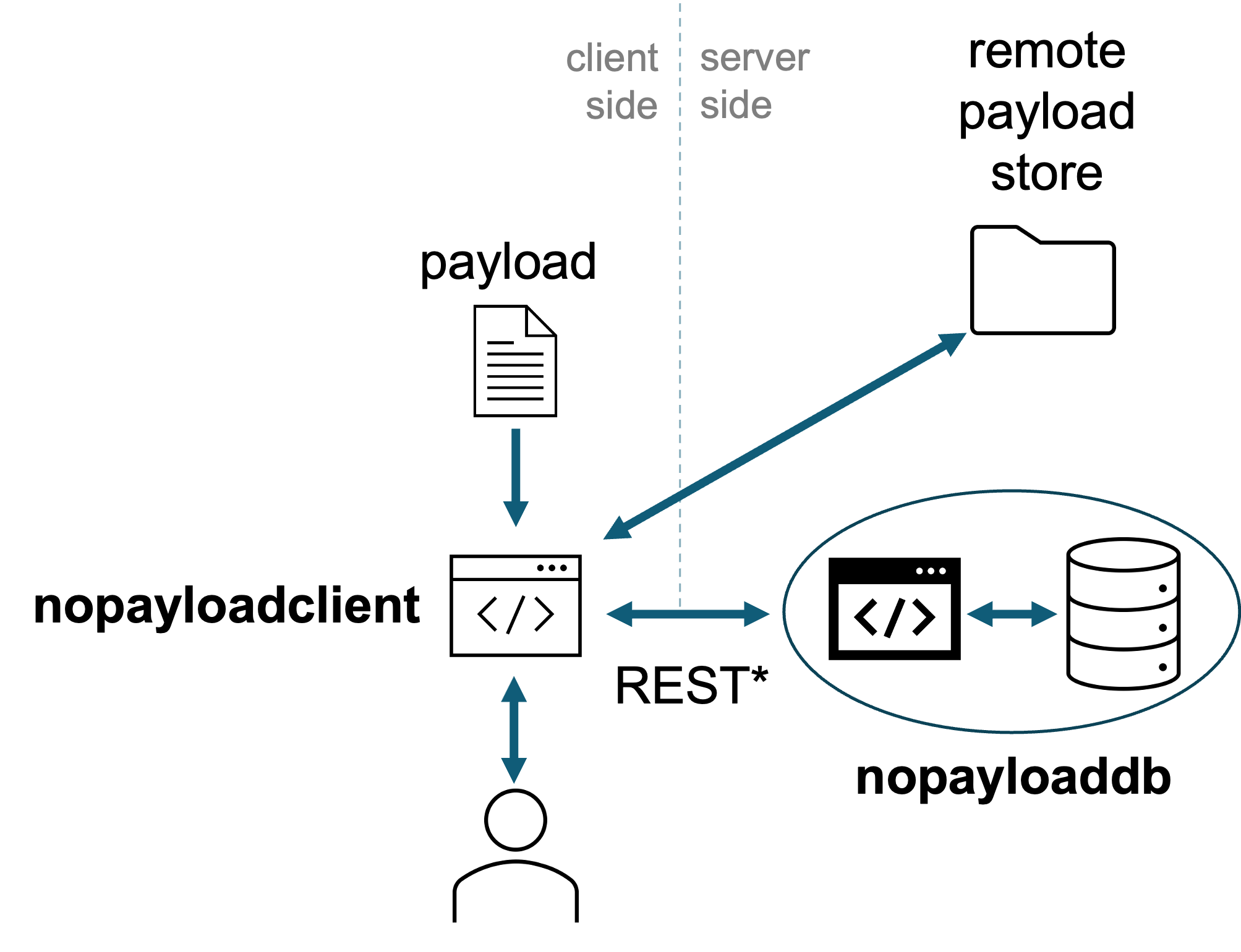}
\caption{Overall layout of the application. The client-side library \textbf{nopayloadclient} communicates with the server-side application, \textbf{nopayloaddb}, through a REST interface. The actual payloads reside in the remote payload storage (e.g. a /cvmfs/ \cite{Buncic:2010zz} file system).}
\label{fig:layout}
\end{figure}

A lightweight C++ client-side library, \textbf{nopayloadclient}, was developed as part of the overall conditions data software suite. Its purpose is to communicate with the server-side service and to handle the payloads while making sure the payload store and the database remain synchronized. The overall layout of the application can be seen in Fig. \ref{fig:layout}.\\
Communication with NoPayloadDB happens through the REST interface, where \textbf{nopayloadclient} basically acts as a wrapper for the cURL library, and provides its own in-memory caching. Among other things, the client allows the following configurations:
\begin{itemize}
    \item The path of the remote payload store can be configured. Read- and write directories are defined separately. The latter can be a list of paths as a failover mechanism in case one of the file systems in unavailable to increase the availability of the service.
    \item The URLs of all payloads for a given type can be overridden locally, so that a user-specified file will be preferred over the ones referenced in the database. This allows to test new calibrations in the context of an existing global tag before publishing them on the system.
    \item The client can be configured not to contact the REST API and instead use a tiny in-memory \textit{fake} database. As the interface and the format of the responses remain exactly the same, this configuration allows developing in an environment where the REST API is not reachable.
\end{itemize}

\subsection{Payload Handling}
When requesting to insert a new payload for a specified global tag, payload type and IoV, the client first calculates the checksum of the payload. Based on the checksum, a path containing subdirectories and a file URL are generated. This avoids duplication of identical payloads and evenly spreads the payloads across (sub-) directories, maximizing the performance of the file system. Furthermore, the checksum can later be used to verify a correct read operation. Next, \textbf{nopayloadclient} checks if the insertion operation would be permitted and if the payload could be copied to its previously determined destination before making the actual insertion request. That way, the synchronization between database and payload store is guaranteed.

\section{Performance Testing}
\label{sec:testing}
As mentioned in Sec. \ref{sec:intro}, very high request rates to the conditions data are expected. Therefore, it is important to systematically test the system's performance in expected conditions. For realistic testing, assumptions must be made about two things: Firstly, the expected database occupancy has to be estimated. Secondly, the performance tests should make requests according to expected access patterns.

\subsection{Database Occupancy Scenarios}
\label{sec:scenarios}
To simulate the database occupancy, \textit{scenarios} were defined. For a given global tag, there are two parameters that can be varied: the number of payload types and the number of IoVs per payload type. The five considered scenarios range from \textit{tiny} with 10 payload types with 10 IoVs each to \textit{worst-case} defined as 200 payload types with 26000 IoVs each.  Table \ref{tab:scenarios} summarizes these scenarios.

\begin{table}[b]
    \centering
    \begin{tabular}{|c|c|c|}\hline
         Scenario &  Payload types & IoVs (per type)\\\hline
         tiny & 10 & 100 (10)\\
         tiny-moderate & 10 & 2000 (200) \\
         moderate & 100 & 20000 (200) \\
         heavy-usage & 100 & 500000 (500) \\
         worst-case & 200 & 2600000 (2600) \\\hline
    \end{tabular}
    \caption{Global tag occupancy scenarios for performance testing.}
    \label{tab:scenarios}
\end{table}

\subsection{Access Patterns}
In real-world usage, many (O(10000)) jobs running on (distributed) computing resources will make requests to the conditions database simultaneously. The vast majority of these jobs will access the data in a read-only fashion. More precisely, they will request the payloads for a given global tag and IoV. Other operations, on the other hand, will be conducted at much lower frequencies (e.g. inserting new payloads, checking the list of all global tags, etc.). For this reason, all performance tests shown in the following run the performance-crucial read query exclusively. Furthermore, in order to minimize the influence of any unwanted hidden caching anywhere in the application, major- and minor IoV are chosen from a uniform random distribution between 0 and the largest possible value. 

\subsection{Methodology}
\begin{figure}
 \centering
 \includegraphics[width=12cm,clip]{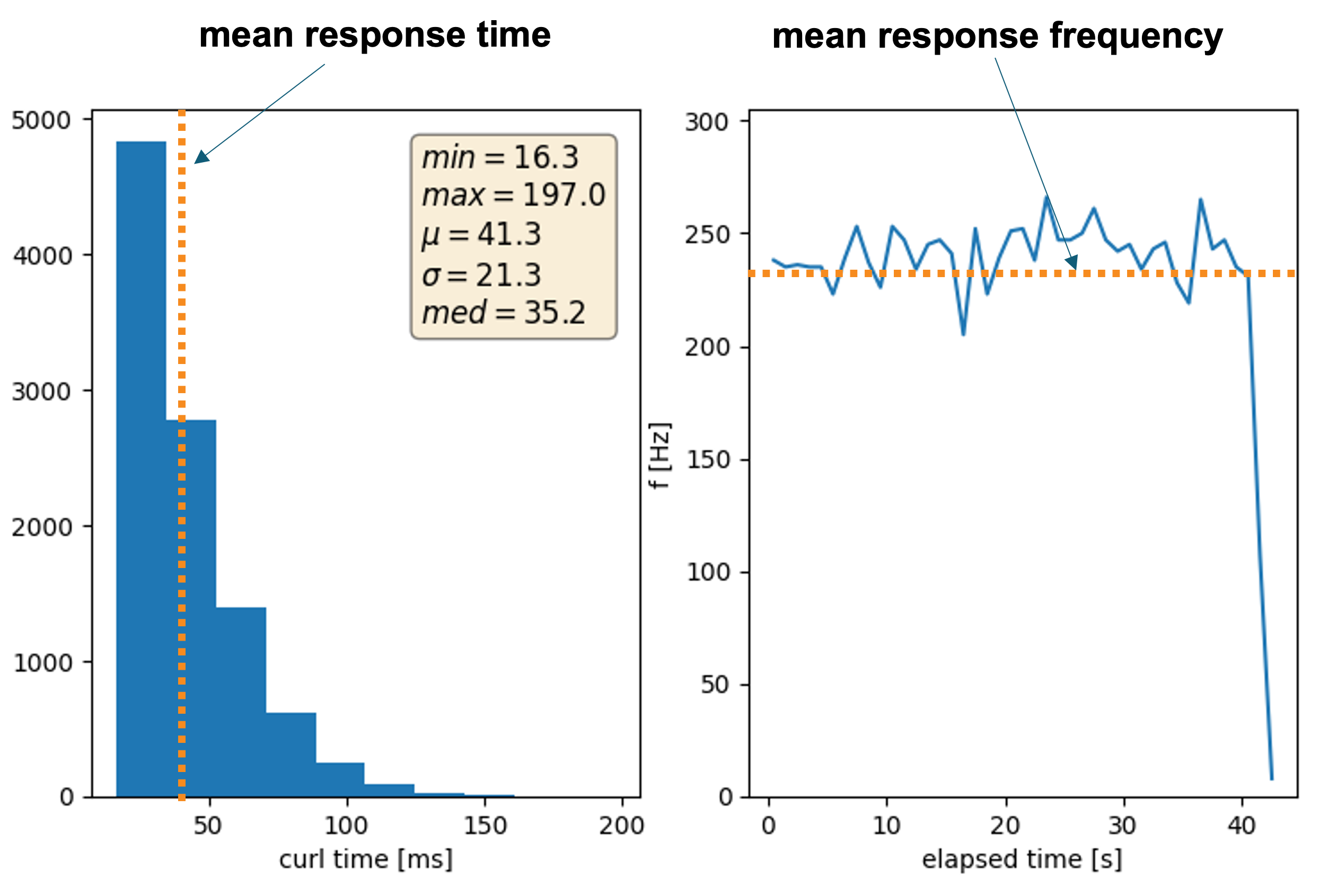}
 \caption{Results of an example test campaign making 10.000 random requests from a single client using asynchronous multithreading. The left shows the distribution of the recorded response times, the right shows the mean response frequency for each second over the duration of the campaign.}
 \label{fig:campaign}
\end{figure}

The performance of the server-side application is evaluated with the following procedure: the client sends a request to the server and saves the corresponding time stamp. When receiving the answer to that request, the client saves the corresponding time stamp as well, resulting in a pair of time stamps from which the response time and request frequency can be calculated. It is worth noting that only the retrieval of the payload URL through the REST API is timed. Retrieving the actual payload from storage is not part of the performance testing. Repeating this procedure in a synchronized fashion using only a single client will result in the request frequency being the inverse of the response time. To generate higher request rates, several concurrent jobs making subsequent requests were run with the help of HTCondor \cite{condor-practice}. The number of concurrent jobs running in parallel was increased until the response frequency flattened out, indicating that a server-side bottleneck was reached - the highest possible response frequency. This testing environment is very similar to the expected real world usage. A second approach was also tested where a single client can generate very high request rates with the help of asynchronous multithreading. In this case, the client proceeds to send requests before receiving answers to the previous ones. It became evident that the asynchronous multithreading approach using a single client generates request frequencies that are high enough to also reach the server-side bottleneck while at the same time offering a significantly shorter turn-around time and less overhead by omitting the batch scheduling and post-processing to collect the results from the various nodes. Therefore, all results shown in the following were derived using the multithreading approach. Also, the server-side application was configured to execute each service component on just a single pod. As an example, Fig. \ref{fig:campaign} shows the results of a test campaign making 10000 random requests.

\subsection{Scaling Results}
\begin{figure}
 \centering
  \includegraphics[width=9.7cm,clip]{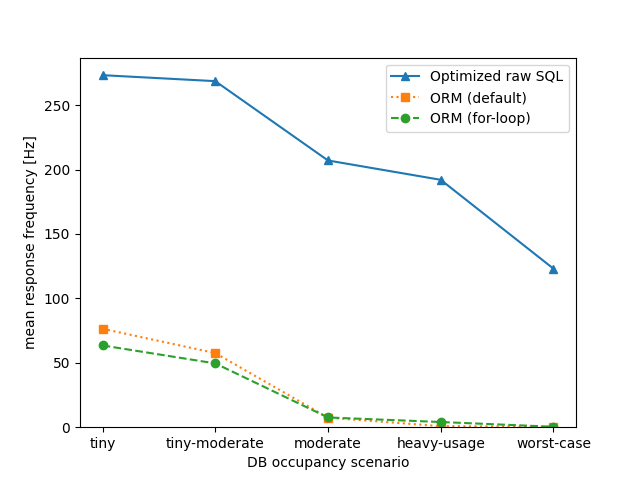}
 \caption{Mean response frequency for 10000 random requests as a function of the database occupancy scenario for three different implementations of the \textbf{payloadiovs} query.}
 \label{fig:endpoints}
\end{figure}
The arrays of response times and -frequencies of a test campaign can be condensed into a few higher-level metrics such as the mean response time and mean request- and response frequency. Conducting several campaigns for different database occupancy scenarios allows to investigate trends in these metrics. Fig. \ref{fig:endpoints} shows this scaling behaviour for three different implementations of the same query. Two of which rely on Django's ORM and exhibit significantly worse performance than the end-point based on an optimized raw SQL query, which was therefore chosen as the default implementation. The mean response frequency remains over 100 Hz even for the worst-case scenario. Additional performance tests were conducted that cannot be covered in detail in this paper. Some of the key results are:
\begin{itemize}
 \item \textbf{Peak request frequency} was simulated by sending 10000 random requests within one second. All requests were successfully answered within one minute, showing that the server-side queuing of requests works as expected and that peaks of request frequencies can be handled with no failures.
 \item \textbf{Scaling with large table sizes} was investigated by inserting additional worst-case global tags until roughly 70 million IoVs resided in the corresponding database table. No significant performance deterioration was observed as expected because of the query being executed as a pure index condition.
 \item \textbf{Insertion and request order} of the IoVs were replaced by different random distributions as well as ascending and descending order. The performance results remained unchanged, suggesting that the database indices work as intended. 
 \item \textbf{Horizontal scaling of each service component} has an effect on the performance. Generally, increasing the number of pods that run the Django application leads to higher response frequencies. However, this scaling is not linear and strongly depends on the considered database occupancy scenario. Horizontal scaling of the nginx service component does not have a significant effect.
\end{itemize}

\section{Conclusion}
\label{sec:conclusion}
The management of conditions data in HEP is a complex endeavor, fraught with challenges such as the heterogeneous structure of the data, granularity, and the need for high access rates. These challenges are similar across various experiments. Based on experience from various collaborations, the HSF published a set of best practices for handling conditions data.\\
Implementing these recommendations, a reference system was developed comprising an easy-to-deploy, scalable server-side application built on PostgreSQL and a Django-REST API. On the client side, a lightweight configurable C++ library facilitates seamless interaction with the server and the handling of the payloads which reside on a distributed file system like CVMFS. Performance tests have shown promising results, with the system capable of consistently handling O(100) Hz request frequency in a worst-case occupancy scenario (spiking up to 10 kHz) without the need for any caching. The number of conditions data entries was scaled up to 70 million without any issues.\\
The practical utility of this framework is evidenced by its successful deployment in the sPHENIX \cite{Campbell_2017} experiment at Brookhaven National Lab (US). Running smoothly on more than 25,000 concurrent jobs, the system has proven itself in a real-world, high-stakes environment. This not only validates the effectiveness of the HSF's recommendations but also establishes the reference implementation as a viable, scalable solution for handling conditions data in various scientific settings.

\bibliography{refs} % Entries are in the refs.bib file

\end{document}